\documentclass[aps,prl,superscriptaddress, twocolumn, footinbib]{revtex4-1}

\usepackage{amsmath}
\usepackage{graphicx}
\usepackage{dcolumn}
\newcolumntype{d}[1]{D{.}{.}{#1}}

\usepackage{bm}
\usepackage{epstopdf}
\usepackage{graphicx}
\usepackage{stmaryrd}
\usepackage{tikz}
\usepackage{lipsum}
\usepackage{pgf}
\usepackage{float}
\usepackage{hyperref}
\usepackage{tabularx}
\usepackage{subfigure}
\newcommand {\bkt} [1] {\langle #1 \rangle}

\definecolor{lightblue}{RGB}{0,170,255}

\newcommand* {\bra}[1]{\langle {#1} |}
\newcommand* {\ket}[1]{| {#1} \rangle}

\newcommand{\appropto}{\mathrel{\vcenter{
			\offinterlineskip\halign{\hfil$##$\cr
				\propto\cr\noalign{\kern2pt}\sim\cr\noalign{\kern-2pt}}}}}

\usepackage[normalem]{ulem}
\newcommand{\dps}{\displaystyle}

\newcommand{\om}{\iffalse}
\newcommand{\pd}[2]{\frac{\partial #1}{\partial #2}}

\newcommand{\ba}{\arraycolsep 0.3ex \begin{array}{rl}}
	\newcommand{\ea}{\end{array}}
\newcommand{\bc}{\begin{cases}}
	\newcommand{\ec}{\end{cases}}
\usepackage{color}
\newcolumntype{C}[1]{>{\centering\arraybackslash}p{#1}}

\begin{document}
	\title{Resonant photovoltaic effect in doped magnetic semiconductors}
	
	\author{Pankaj Bhalla}
	\affiliation{Beijing Computational Science Research Center, Beijing 100193, China}
	\author{Allan H. MacDonald}
	\affiliation{Department of Physics, The University of Texas at Austin, Austin Texas 78712, USA}
	\author{Dimitrie Culcer}
	\affiliation{School of Physics, University of New South Wales, Sydney 2052, Australia}
	\affiliation{ARC Centre of Excellence in Future Low-Energy Electronics Technologies, UNSW Node, Sydney 2052, Australia}

\date{\today}

\begin{abstract}
The rectified non-linear response of a clean undoped semiconductor to an {\em ac} electric field includes a well known intrinsic contribution -- the \textit{shift current}. We show that when Kramers degeneracy is broken, a distinct second order rectified response appears that is due to  Bloch state anomalous velocities in a system with an oscillating Fermi surface. This effect, which we refer to as the resonant photovoltaic effect (RPE), 
produces a resonant galvanic current peak at the interband absorption threshold in doped semiconductors or semimetals with approximate particle-hole symmetry. We evaluate the RPE for a model of the surface states of a magnetized topological insulator.
\end{abstract}

\maketitle

{\em Introduction:}---
The interband coherence responses of crystals to {\em dc} and {\em ac} driving electric fields 
have both been studied extensively in recent years.
For  example, researchers have come to appreciate that the intrinsic 
anomalous velocity {\em dc} response, which is due to interband coherence
and related to momentum-space Berry curvature, is essential for the chiral anomaly
\cite{Shuang_NatMat2016, Burkov_PRB2012} in Weyl semimetals, and that it often dominates the anomalous 
quantum Hall effect of magnetic materials.\cite{nagaosa_RMP2010, Liu_AnnRev2016} 
Separately a number of conceptually novel non-linear response effects \cite{Nagaosa_NRR_NC2018} 
have been identified recently that involve inter-band coherence.
Notably, the non-linear optical response of a semiconductor at frequencies above the band gap  
includes an intrinsic {\em dc} photocurrent associated with 
an interband-coherence related shift of intra-cell coordinates.  
The intrinsic shift current \cite{Belinicher_SovPhy1978, Kraut_PRB1981, Ivchenko_SovPhy1984, Lyanda_JETPL1987, Belinicher_Ferro1988, Fridkin_book, Sipe_PRB2000, Shengyuan_PRB2009, Sipe_PRB2010, Sun_NL2010, ivanov_PRB2011, Andrew_PRL2012, rappe_16, alexey_16, shilie_16, Moore_NATComm2017, Kim_Shift_PRB2017, Hamamoto_NLSC_PRB2017, wenhui_17, meier_17, tokura_17, nagaosa_18, Rostami_NLAH_PRB18, Isobe_2018, Golub_JETP2017, Golub_PRB2018, Tarasenko_ADP2019} effect has 
received particular attention because it is is closely related to topological band 
characteristics \cite{Inti_PRL2015, nagaosa_16, nagaosa_16a}, and has been identified experimentally
in some non-centrosymmetric ferroelectrics \cite{dawson_07, rappe_12, woerner_14, meier_16}.
In this Letter we identify a new non-linear response effect by showing that the {\em dc} galvanic photocurrent 
response of doped semiconductors can contain an anomalous velocity contribution.

\begin{figure}[htbp]
	\centering
	\subfigure[noonleline][]
	{\label{fig: expt}\includegraphics[height=4cm,width=4.5cm]{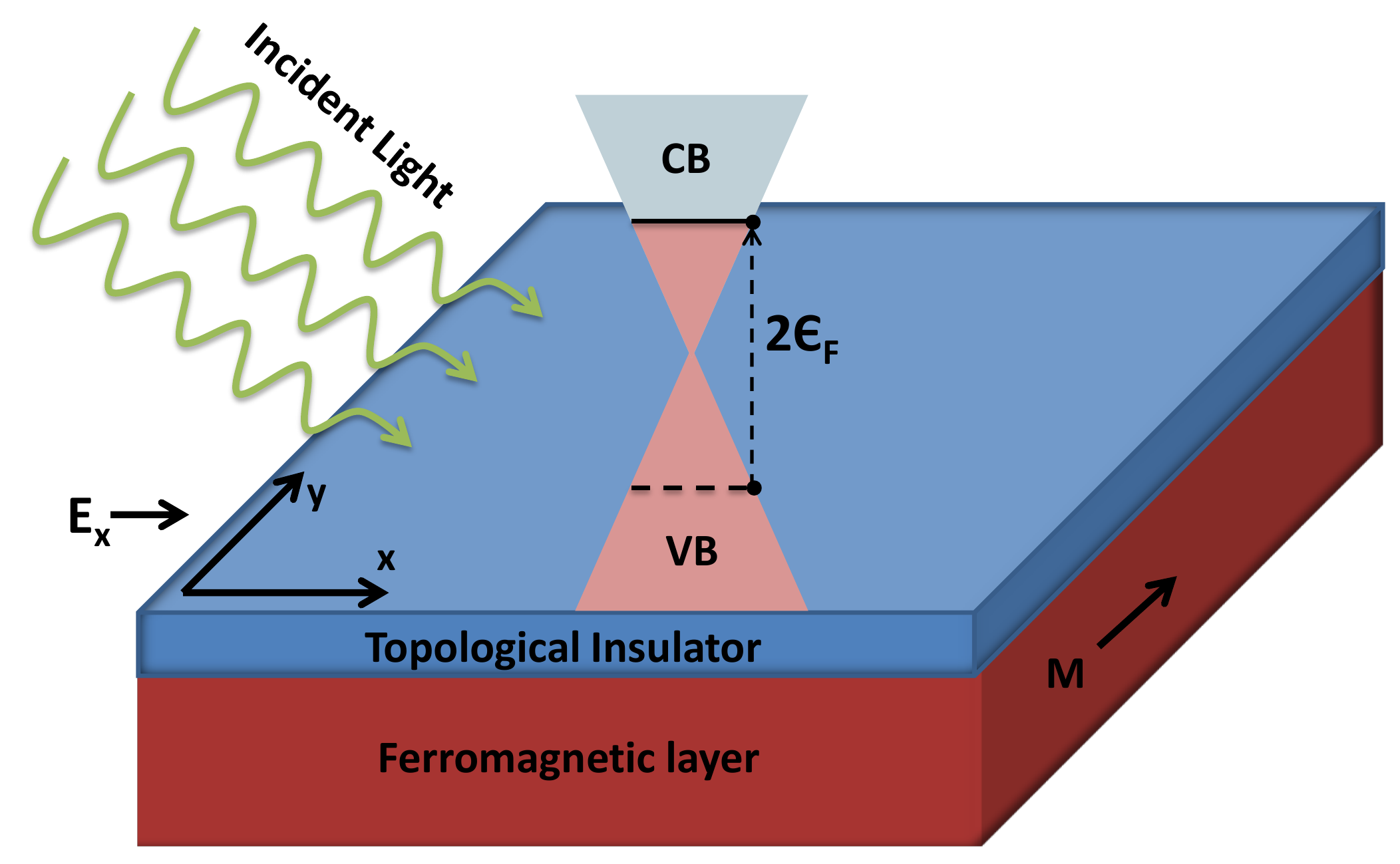}}
	\subfigure[noonleline][]
	{\label{fig: expt2}\includegraphics[height=4cm,width=4cm]{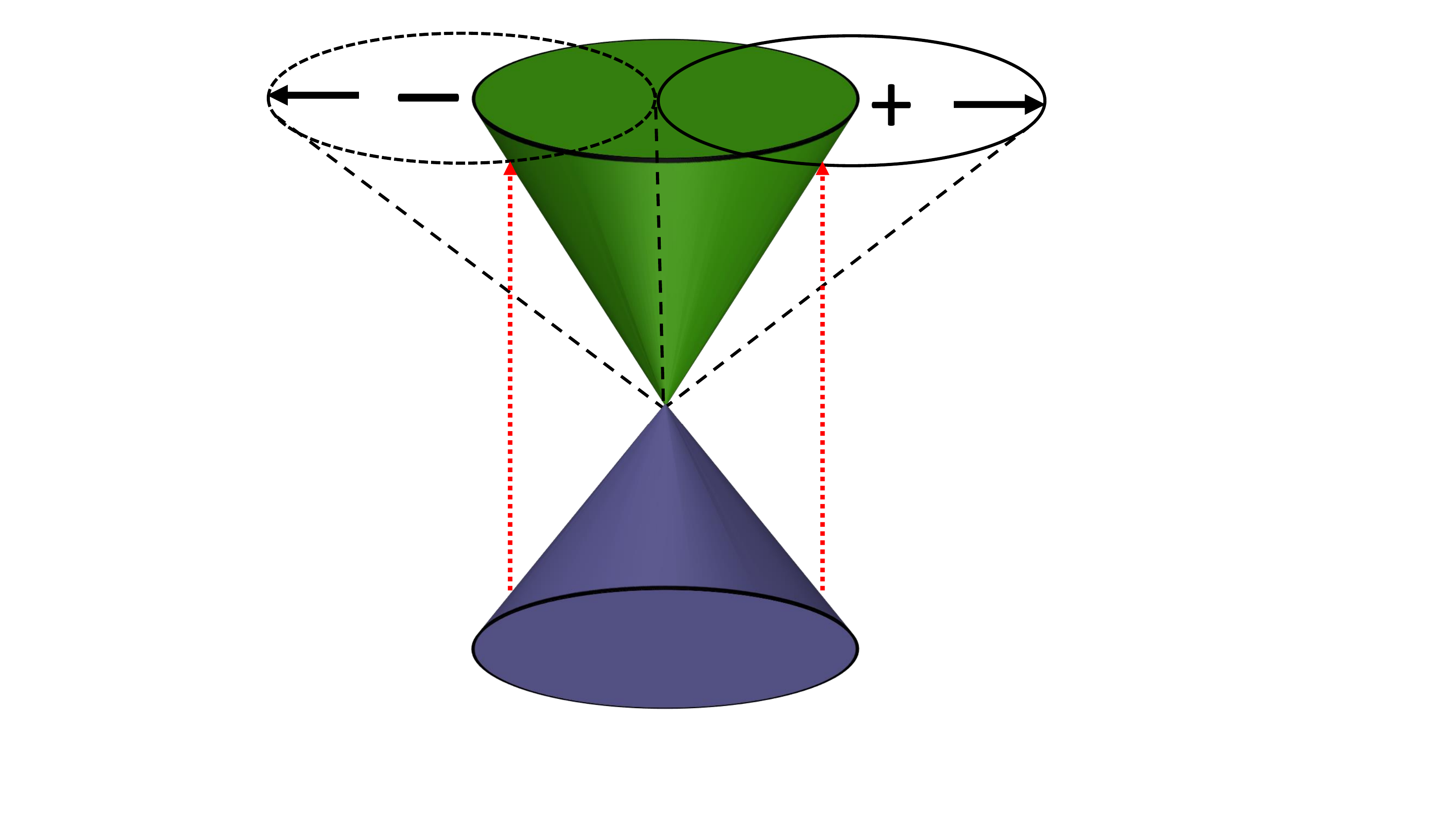}}
	\subfigure[noonleline][]
	{\label{fig: current}\includegraphics[height=4cm,width=6cm]{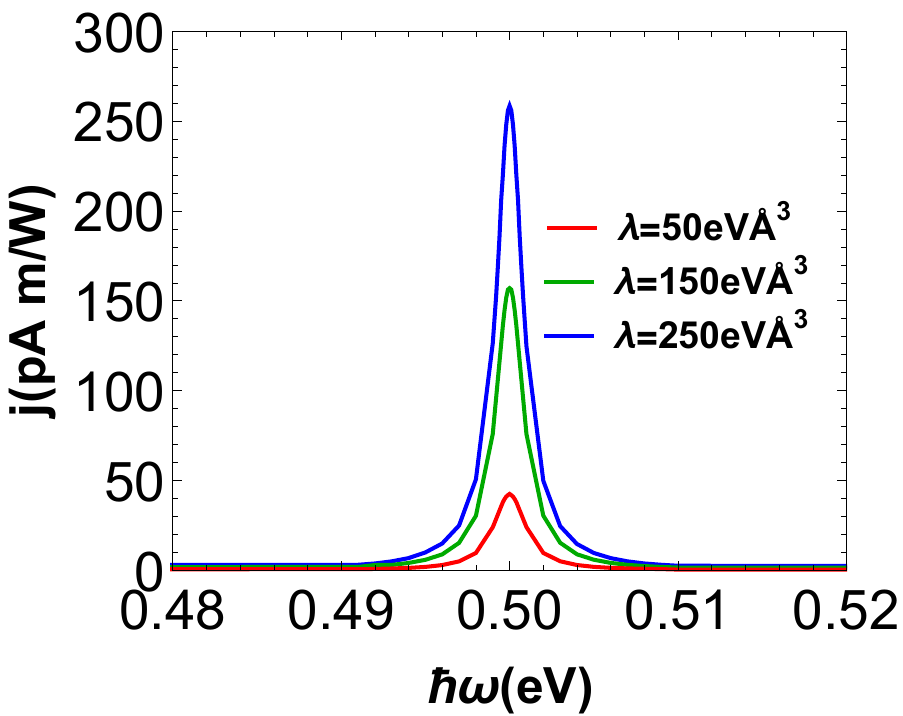}}
\vspace{-0.2cm}
	\caption{(a). Resonant photovoltaic effect current induced by linearly polarized light incident 
	on the surface of a warped topological insulator placed on a ferromagnetic layer with an in-plane magnetization. (b). Carriers are excited from the valence band to the oscillating Fermi surface. (c). RPE response of magnetized topological insulator surface states with different values for the warping coefficient $\lambda$ using the model parameters $\varepsilon_F = 250$ meV, $A = 2.55$ eV\AA, $T = 1K$, $M = 10$ meV, and $\tau = 1$ps explained in the text. The blue curve corresponds to the experimental value of $\lambda$ for Bi$_2$Te$_3$.}
	\label{figure1}
\end{figure}

The understanding of inter-band coherence and its relation to disorder in the non-linear optical response
of semiconductors is still in its infancy. Most studies to date have focused on undoped materials, although possible Fermi surface effects in doped systems have started to gain attention \cite{Isobe_2018, Du_2018, Facio_PRL2018, Konig_PRB2019, Nandy_Tilt2019} very recently. The resonant photovoltaic effect (RPE) mechanism for rectified response to linearly polarized light is due to the combination of Bloch state anomalous velocities and Fermi surface shifts, which both oscillate when driven by an {\em ac} field and produce a current with a non-zero time average. The RPE involves an interplay between Bloch state wave function topology, disorder, and inter-band optical excitation. The RPE is active in doped semiconductors with with broken time-reversal symmetry, and strongest in semiconductors with approximate particle-hole symmetry, as illustrated in Fig.~\ref{fig: current}. It is therefore especially strong in magnetized topological materials whose surface states have approximate particle-hole symmetry, reflecting the fundamental connection between non-linear response and non-trivial band topology \cite{Inti_PRL2015, xiaoqin_NATComm2015, You_BCD_TMD_PRB2018, Binghai_BCD_PRB18, BCD_2018, Pilidi_IEEE2019}, and the importance of the Berry curvature in non-linear optical response \cite{Hosur_PRB2011}. The RPE is related in part to the non-linear Hall conductivity, which contains a related intrinsic contribution proportional to the Berry curvature dipole \cite{Inti_PRL2015, You_BCD_TMD_PRB2018, Binghai_BCD_PRB18, BCD_2018} but may also have extrinsic contributions \cite{Gao_Shift_PRL2014, Gaoyang_PRB2018, Isobe_2018, Nandy_Tilt2019}. Non-linear phenomena in topological materials have been discussed previously e.g. the observation of the non-linear Hall effect \cite{NLH_2019, Kang_NM2019}, the prediction of a non-linear anomalous Hall effect \cite{NLH_Tune_2DM2018}, and valley-driven second harmonic generation \cite{Hipolito_2DM2017}.

\begin{figure}[tbp]
	\centering
	\includegraphics[height=6cm,width=8cm]{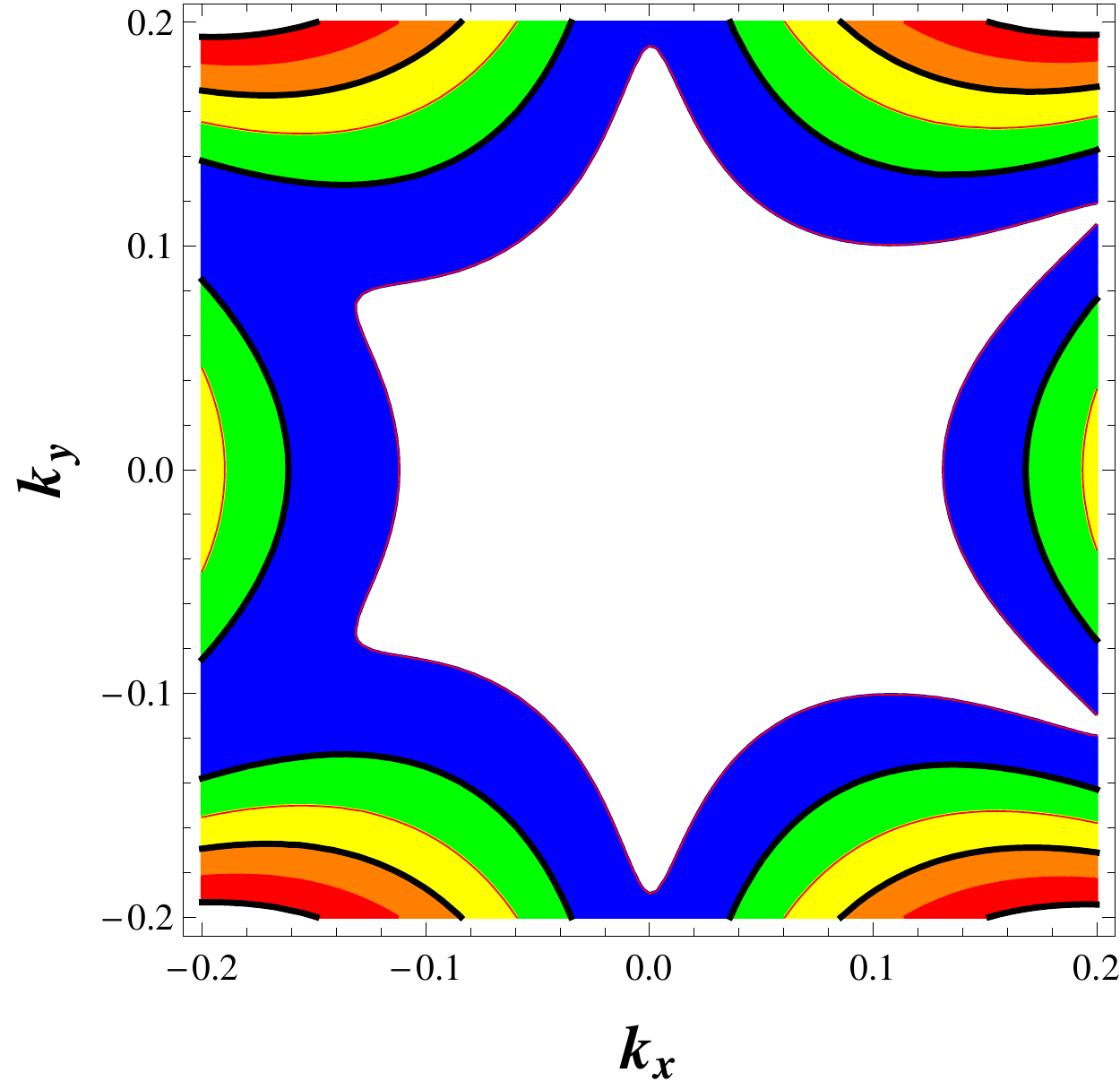}
	\caption{Constant energy contour showing the breaking of Kramers degeneracy by the in-plane magnetization. 
	For this figure the warping constant $\lambda=200$eV\AA$^3$, while the magnetic exchange energy $M=0.05$eV.}
	\label{contour}
\end{figure}

{\em Theory of the Resonant Photovoltaic Effect:}--- 
We now outline the transport theory that we use to identify and evaluate the RPE; a detailed derivation is provided in the supplementary material. We consider a Hamiltonian of the general form $\hat{H} = \hat{H}_0 + \hat{H}_E + \hat{U}$, where $\hat{H}_0$ is a crystal Bloch-state Hamiltonian, $\hat{H}_E (t) = e{\bm E} (t) \cdot \hat{{\bm r}}$ the interaction with a time-dependent external electric field that is assumed spatially uniform, and $\hat{U}$ is the random disorder potential. The impurities are uncorrelated and the average of $|\bra{{\bm k}m}\hat U\ket{{\bm k}'m'}|^2$ over impurity configurations is $n_i |\bar{U}_{{\bm k}{\bm k}'}^{mm'}|^2/V$, where $n_i$ is the impurity density, $V$ the crystal volume and $\bar{U}_{{\bm k}{\bm k}'}^{mm'}$ the matrix element of the potential of a single impurity. We consider short-range impurities such that $U(\boldsymbol{r})=U_0\sum_i \delta(\boldsymbol{r}-\boldsymbol{r_i})$, with $\boldsymbol{r_i}$ labeling impurity sites. We focus on temperatures close to absolute zero, so that phonon scattering is negligible. The system is described by a density operator $\hat{\rho}$, which obeys the quantum Liouville equation, as described in \cite{culcer_17}. The quantum kinetic equation for $\langle\rho\rangle$, the density matrix averaged over impurity configurations, reads:
\begin{equation}
\frac{d\langle\rho\rangle}{dt} + \frac{i}{\hbar}[H_0, \langle \rho\rangle] + J(\langle \rho \rangle) = -\frac{i}{\hbar}[H_{E},\langle\rho \rangle].
\label{QKE}
\end{equation}
In the Born approximation, the scattering term \cite{culcer_17}
\begin{equation}\label{J}
J(\bkt{\rho}) = \frac{1}{\hbar^2} \, \int^\infty_0 dt' \, \bkt{[U, [e^{-\frac{iH_0t'}{\hbar}}U e^{\frac{iH_0t'}{\hbar}}, \bkt{\rho(t)}]]}.
\end{equation}
The impurity average restores translational periodicity so that in the crystal momentum representation $\langle\rho\rangle$ remains diagonal in the wave vector ${\bm k}$. We expand the density matrix in powers of the electric field as $\langle\rho\rangle = \langle\rho\rangle^{(0)} + \langle\rho\rangle^{(1)} + \langle\rho\rangle^{(2)} + ...$ where the superscript $^{(n)}$ refers to order $n$ in the electric field. The equilibrium part $\langle\rho\rangle^{(0)}$ is the solution of Eq.~(\ref{QKE}) with the RHS set to zero. It is diagonal in the band index $m$ and has the form $\langle\rho\rangle^{(0)}_{mm'} = f^0_{m{\bm k}} \delta_{mm'}$, where $f^0_{m{\bm k}} \equiv f^0_{m{\bm k}} (\varepsilon_{m{\bm k}})$ is the Fermi-Dirac distribution occupation probability at the energy $\varepsilon_{m{\bm k}}$ of band $m$. To evaluate $\langle\rho\rangle^{(1)}$ we set $\langle\rho\rangle \rightarrow \langle\rho\rangle^{(1)}$ on the LHS of Eq.~(\ref{QKE}), and $\langle\rho\rangle \rightarrow \langle\rho\rangle^{(0)}$ on the RHS. Finally, $\langle\rho\rangle^{(2)}$ contains the \textit{non-linear} response of second order in the electric field, which is of interest to us in this work. To determine it we set $\langle\rho\rangle \rightarrow \langle\rho\rangle^{(2)}$ on the LHS of Eq.~(\ref{QKE}), and $\langle\rho\rangle \rightarrow \langle\rho\rangle^{(1)}$ on the RHS. 

Because of the important role of the commutator $[H_0, \langle \rho\rangle]$, which accounts for dynamics in the absence of electric fields and disorder, it is convenient to make the decomposition $\langle\rho\rangle_{\bm k} = f_{\text{d},{\bm k}} + f_{\text{od},{\bm k}}$ with $f_{\text{d},{\bm k}}$ and $f_{\text{od}, {\bm k}}$ respectively purely diagonal and purely off-diagonal in the band indices. The diagonal response $f_{\text{d},{\bm k}}$ tracks Bloch state repopulation while the off-diagonal part $f_{\text{od},{\bm k}}$ accounts for interband coherence. These two responses can be expanded separately in powers of electric field as $f_{\text{d},{\bm k}} = f^{(0)}_{\text{d},{\bm k}} + f^{(1)}_{\text{d},{\bm k}} + f^{(2)}_{\text{d},{\bm k}}+\cdots$ and $f_{\text{od},{\bm k}}=f^{(1)}_{\text{od},{\bm k}}+f^{(2)}_{\text{od},{\bm k}}+\cdots$. The zeroth order term in the expansion is the equilibrium term $f^0_{m{\bm k}} (\varepsilon_{m{\bm k}})$ introduced above, which is diagonal in the Bloch eigenstate representation, hence $f_{\text{d},{\bm k}}$ starts at zeroth order in the electric field while $f_{\text{od},{\bm k}}$ starts at first order in the electric field. It is useful to separate the quantum kinetic equation Eq.~\ref{QKE} into coupled equations for $f_{\text{d},{\bm k}}$ and $f_{\text{od},{\bm k}}$. The scattering term is linear in density matrix and couples the diagonal and off-diagonal response: $J(\langle\rho\rangle) = J_{\text{d}}(f_{\text{d},{\bm k}} + f_{\text{od},{\bm k}}) + J_{\text{od}}(f_{\text{d},{\bm k}} + f_{\text{od},{\bm k}})$. To determine $J_{\text{od}}(f_{\text{d},{\bm k}})$, first $f_{\text{d},{\bm k}}$ is found, then it is fed into Eq.\ (\ref{J}), and the off-diagonal part is selected. 

To illustrate the RPE we consider linearly polarized light ${\bm E}(t) = {\bm E} \, \cos \omega t$. The electric field and scattering terms both connect $f_{\text{d},{\bm k}}$ and $f_{\text{od},{\bm k}}$. The solution in powers of ${\bm E}(t)$ is:
\begin{equation}
\arraycolsep 0.3 ex
\begin{array}{rl}
\displaystyle \frac{df^{(n)}_{\text{d},{\bm k}}}{dt} + J_{\text{d}}[f^{(n)}_{\text{d},{\bm k}}] = \displaystyle \frac{e{\bm E}(t)}{\hbar} \cdot \frac{\partial f^{(n - 1)}_{\text{d},{\bm k}}} {\partial {\bm k}} & - J_{\text{d}}[f^{(n)}_{\text{od},{\bm k}}] \\ [2ex]
\displaystyle \frac{df^{(n)}_{\text{od},{\bm k}}}{dt} + \frac{i}{\hbar} \, [H_{0{\bm k}}, f^{(n)}_{\text{od},{\bm k}}] + J_{\text{od}}[f^{(n)}_{\text{od},{\bm k}}] & = \displaystyle \frac{e{\bm E}(t)}{\hbar} \cdot \frac{D f^{(n - 1)}_{\text{od},{\bm k}}} {D {\bm k}} \\ [2ex]
& \displaystyle - J_{\text{od}}[f^{(n)}_{\text{d},{\bm k}}].
\label{Bloch_QKE}
\end{array}
\end{equation}
The covariant derivative $ \frac{D f_{\text{od},{\bm k}}} {D {\bm k}} = \frac{\partial f_{\text{od},{\bm k}}} {\partial {\bm k}} - i \, [ \mathcal{\bm R}_{\bm k}, f_{\text{od},{\bm k}}] $ arises from the ${\bm k}$-dependence of the basis functions. The Berry connection $\mathcal{\bm R}_{\bm k}^{mm'} = \langle u_{\bm k}^{m}\vert i \frac{\partial u_{\bm k'}^{m'}}{\partial {\bm k}}\rangle$, with $\vert u^m_{\bm k} \rangle $ the lattice-periodic Bloch function. The covariant derivative term is absent in the equation for $f_{\text{d},{\bm k}}$ because the commutator has no diagonal terms. It also appears in the current density operator ${\bm j} = -(e/\hbar) \, D H_{0{\bm k}}/D{\bm k}$. We use this approach to evaluate non-linear response in the limit $\varepsilon_F\tau/\hbar \gg 1$.

The general solution of Eq.~\ref{QKE} up to second-order in the electric field is derived in the Supplement. The linear response contains the oscillatory factors $e^{\pm i\omega t}$ as required for time-independent unperturbed Hamiltonians. The second order response has both second-harmonic term $ \propto e^{\pm 2i\omega t}$, and the time-independent terms on which we focus in which the $e^{\pm i\omega t}$ factors cancel. In the strong disorder $\omega \tau \ll 1$ limit a clear hierarchy can be established in powers of the impurity density as explained in detail in \cite{culcer_17} and can be straightforwardly extended to non-linear response, as done in part in \cite{Nandy_Tilt2019}. In the weak disorder limit $\omega \tau \gg 1$, one naively expects scattering to play virtually no role. This is because, firstly, the \textit{cross-scattering terms} $J_{\text{d}}(f_{\text{od},{\bm k}})$ and $J_{\text{od}}(f_{\text{d},{\bm k}})$ connecting $f_{\text{d},{\bm k}}$ and $f_{\text{od},{\bm k}}$ are suppressed by $1/\omega \tau$ and higher powers. Hence it appears that $f_{\text{d},{\bm k}}$ and $f_{\text{od},{\bm k}}$ can be treated independently. Secondly, the leading term in $f_{\text{d},{\bm k}}$ simply yields the Drude conductivity, which at high frequencies is $\propto 1/\omega$. We specialize here to frequencies where $\omega \tau \gg 1$ and focus on systems in which the second harmonic terms are suppressed by factors $\propto (\omega \tau)^{-1}$ and higher, as is the case in Bi$_2$Te$_3$ with an in-plane magnetization, introduced above in Fig.~\ref{figure1}. 

The RPE arises primarily from the second-order off-diagonal response driven by the first order diagonal response.  In the limit $\omega \tau \gg 1$, it is easy to show that the first order diagonal response is 
$f_{\text{d},{\bm k}}^{1} = \partial f^0/\partial k_x (e E_x/\omega) \sin(\omega t)$. The set of ${\bm k}$-vectors that are occupied oscillates in ${\bm k}$-space in a manner that is out-of-phase with the electric field and does not contribute to dissipation. At the same time it follows from the second of Eqs.(3) that at finite frequencies the response of a given occupied ${\bm k}$ to the oscillating electric field also contains a piece that is out-of-phase, and is resonant at the interband transition energy. These two out-of phase resspones combine to yield a current in the electric field direction that has a non-zero time average. The out-of phase current from the inter-band coherence response is a partner of the in-phase anomalous velocity response that explains the quantum anomalous Hall effect in many materials, and remains finite in the dc limit. Combining these two effects we find that, in the second-order response there is one, and only one, term in $f_{\text{od},{\bm k}}$ that is responsible for the peak in the RPE current, Fig.~\ref{fig: current}. For a two-band system with particle-hole symmetry the band index $m \in \{ +, - \}$, as is the case for Bi$_2$Te$_3$ in Eq.~(\ref{Bi2Te3}) considered below, and $\varepsilon_{\bm k}^- = - \varepsilon_{\bm k}^+$, this yields the first contribution to the RPE current
\begin{equation} \label{jod}
j_{x,\text{od}}^{(2)} = \frac{e^3E_x^2}{4\hbar}\int \frac{d^2k}{(2\pi)^2} \, \frac{\vert\mathcal{R}_{k_x}^{+-}\vert^2 (\partial f^0_{+{\bm k}}/\partial k_x) (\hbar/\tau)}{(2\varepsilon_{k}^+ -\hbar\omega)^2+(\hbar/\tau)^2}.
\end{equation}
As $T \rightarrow 0$ the derivative of the Fermi function tends to $-\delta(\varepsilon_F - \varepsilon_{\bm k}^+)$, so the RPE current becomes a Lorentzian centered around $\hbar \omega = 2\varepsilon_F$, as expected from Fig.~\ref{fig: current}. If we examine the value at the peak itself, setting $\hbar \omega = 2\varepsilon_F$ in the integrand, it is immediately seen that the integrand is $\propto \tau$, and $(\partial f^0_{+{\bm k}}/\partial k_x) \, \tau$ is the displacement of the Fermi surface. Noting that $\partial f^0_{+{\bm k}}/\partial k_x = (\partial f^0_{+{\bm k}}/\partial \varepsilon_{\bm k}) \, (\partial \varepsilon_{\bm k}/\partial k_x)$, it is clear that $2 \, (\partial \varepsilon_{\bm k}^+/\partial k_x) \, \tau$ corresponds to the displacement undergone by a particle excited from a state in the valence band with group velocity $- (1/\hbar) (\partial \varepsilon_{\bm k}^+/\partial k_x)$ to a state in the conduction band with group velocity $+ (1/\hbar) (\partial \varepsilon_{\bm k}^+/\partial k_x)$. Evidently, if Kramers degeneracy is present, so that $\varepsilon_{\bm k}^+ = \varepsilon_{- \bm k}^+$, the displacements cancel between opposite sides of the Fermi surface. So Kramers degeneracy needs to be broken for the effect to be finite, which in Bi$_2$Te$_3$ is accomplished by the in-plane magnetic field, as illustrated in Fig.~\ref{contour}. 

An additional contribution of the same order in the electric field and scattering strength $n_i$ arises in our formalism by taking $f_{od,{\bm k}}^{(2)}$, which leads to Eq.~(\ref{jod}), feeding it into the scattering term, and taking the diagonal part, which acts as a driving term for $f_{d,{\bm k}}^{(2)}$ as follows,
\begin{equation}
\frac{df^{(2)}_{\text{d},{\bm k}}}{dt} + J_d[f_{d,{\bm k}}^{(2)}] = - J_d[f_{od,{\bm k}}^{(2)}],
\end{equation}
yielding an additional term in the RPE current
\begin{equation}
j_{x,\text{d}}^{(2)} = \frac{e\tau}{\hbar} \int \frac{d^2k}{(2\pi)^2} \,  \pd{\varepsilon_{\bm k}^+}{k_x} J_d[f_{od,k}^{(2)}].
\end{equation}
This corresponds to inter-band transitions driven by scattering and demonstrates that, contrary to naive expectation, scattering plays a crucial role in the DC current, as do the \textit{cross-scattering terms}. At higher temperatures phonon scattering must be taken into account. The complicated many-body terms that come in through the Pauli blocking factors will be considered in a future study. Likewise, our present study does not incorporate many-body interactions, which may alter the effect at a quantitative level as in linear response.

{\em Resonant photovoltaic effect for Warped Topological Insulator Surface States:}---
Topological insulators such as Bi$_2$Te$_3$ can host strong spin-orbit torques \cite{yasuda_PRL2017, Yang_NatPhy2018, BMER_2019}, and 
produce strong spin-orbit coupling signatures in optics, transport and 
magnetism \cite{Zhang_PhyTod2010, Kane_RMP2010, Macdonald_PRL2010, Cortijio_PRL2010, Gambardella_NatPhy2015, Sankar_SciRep2016, Gambardella_Nat2011, Buhrman_Sci2012, Emori_NatMat2013, Nikolic_SOT, Manchon_SOT, Nikolic_Co_NL2017, Xiao_SOT_PRB2017, Ado_SOT_PRB2017, Mellnik_Nature2014, MacNeill_NP2017, Li_WTe2_NC2018, Wang_NC2017, Han_PRL2017, Khang_NM2018}. 
Time-reversal symmetry breaking 
in these systems can be accomplished by placing the topological insulator on a ferromagnet, as sketched in Fig.~\ref{contour}. A sizable proximity effect can lead surface-state exchange fields parallel to the magnetization of order 10 meV \cite{Luo_PRB2013, Eremeev_PRB2013}. 
The surface state Hamiltonian 
$H_0 = H_R + H_M + H_W$, where  $H_R = A (\sigma_x k_y - \sigma_y k_x)$ is the Rashba spin-orbit interaction with $A$ a material-specific constant, and the $\sigma_i$'s are Pauli matrices. 
The exchange term $H_M = \boldsymbol{\sigma}\cdot \boldsymbol{M}$ with $\bm{M} \parallel \hat{\bm y}$.  We will consider non-linear response to an electric field ${\bm E} \parallel \hat{\bm x}$. 
The warping term $H_W = \lambda \sigma_z (k_x^3 - 3 k_x k_y^2)$ describes hexagonal warping 
that causes the Fermi surface to acquire its well-known snowflake 
shape \cite{Fu_PRL2009, Murkani_PRB2011, Chang_PRB2015, Akzyanov_PRB2018}. The quasiparticle energy dispersion for this model Hamiltonian is
particle-hole symmetric with 
\begin{equation}
\label{Bi2Te3}
\varepsilon_{\bm k}^\pm=\pm\sqrt{A^2 k^2 + M^2 + 2A k M \cos\theta + \lambda^2 k^6 \cos^23\theta},
\end{equation}
where $\theta= \arctan(k_y/k_x)$ is the polar angle of the wave-vector ${\bm k}$. In Fig.~\ref{fig: current} we have plotted the total RPE current as a function of photon 
energy $\hbar\omega$ at different warping constants and at $\varepsilon_F = 250$meV. 
For direct comparison, we use the same pA/m units for the current density as in Ref.~\cite{Kim_Shift_PRB2017}. 
The intrinsic shift current as calculated in Ref.~\cite{Kim_Shift_PRB2017} is zero in this configuration, thus the entire signal is from the RPE current. The RPE current has a sharp and tunable peak at $\hbar \omega = 2\varepsilon_F$,
an attractive feature for potential photovoltaics applications. 
We list estimated values for the peak RPE current in Table~\ref{tab:peak} for a series of Fermi energies
$\varepsilon_F$ smaller than the bulk gap $\sim 300$meV.

\begin{table}[tbp]
	\centering
	\caption{Peak resonant photovoltaic effect current in pA m/W partitioned into band diagonal and off-diagonal contributions 
	at different Fermi energies for magnetic exchange energy 
	M$=10$meV \cite{Luo_PRB2013}, $\lambda=250$eV\AA$^3$ \cite{Fu_PRL2009}, $\tau=1$ps,  A$=2.55$eV\AA \cite{Fu_PRL2009} and T$=1$K. The total current is in the right column.}
	\label{tab:peak}
	\begin{tabular}{|C{1.8cm}|C{1.8cm}|C{1.8cm}|C{2.6cm}|}
		\hline
		$\epsilon_F$ (meV) & $j_{x,od}^{(2)}$ & $j_{x,d}^{(2)}$ & $j_{x}^{(2)}=j_{x,d}^{(2)}+j_{x,od}^{(2)}$   \\
		\hline	
		50	 &  3.8& 4.7&  8.5 \\
		100   &  21&  14&   35 \\
		150	 &  57&  29&   86 \\
		200   &  99&  62&   161 \\
		250   &  140&  119& 259 \\
		300    &  180&  142& 322 \\
		\hline
	\end{tabular}
	
\end{table}

{\em Discussion:}--- The physical explanation of the RPE is as follows. For $\hbar\omega \ll 2\varepsilon_F$ no carriers can be excited into the conduction band. As $\hbar\omega$ approaches $2\varepsilon_F$ electrons can be excited from energy $-\varepsilon_F$ in the valence band to $\varepsilon_F$ in the conduction band. The constant energy surface at $-\varepsilon_F$ in the valence band is \textit{not} oscillating, while the Fermi surface $\varepsilon_F$ in the conduction band oscillates under the action of the $ac$ electric field. Importantly, the Fermi surface is inversion asymmetric due to the breaking of Kramers degeneracy by the in-plane magnetization. Hence, as the Fermi surface oscillates, its displacement along $+k_x$ is not equal to that along $-k_x$, resulting in a net current. This current depends on the anomalous velocity, contained in the Berry connection, and on the momentum relaxation time $\tau$. This effect occurs only for excitation around the Fermi surface, which explains the resonance in the signal. More importantly, if time reversal symmetry is preserved, the system possesses Kramers degeneracy, meaning that $\varepsilon_{\bm k} = \varepsilon_{-{\bm k}}$, and the effect cancels between the two sides of the Fermi surface as the electric field oscillates along the $\hat{\bm x}$-axis. In the example given the in-plane magnetization breaks Kramers degeneracy, which can be seen clearly in Fig.~\ref{contour}, so that the positive and negative $\hat{\bm x}$-axes are not equivalent, and the effect does not cancel as the electric field oscillates in the positive and negative $\hat{\bm x}$-directions. The contributions from the off-diagonal and diagonal parts of the density matrix are comparable in magnitude.

\begin{figure}[tbp]
	\centering
	\includegraphics[width=8cm,height=6cm]{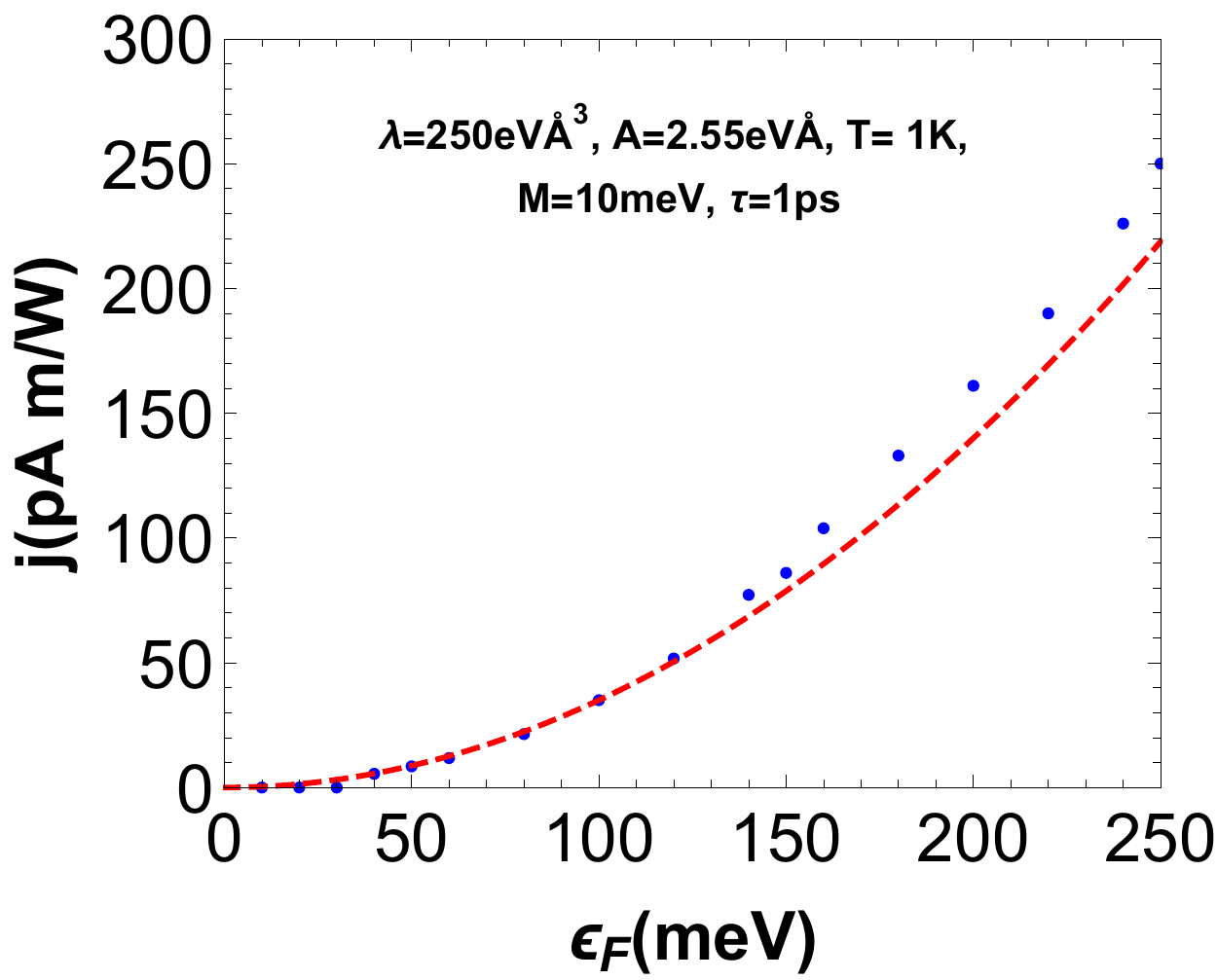}
	\caption{Resonant photovoltaic effect current as a function of Fermi energy. The blue dotted curve corresponds to the peak value of the RPE current as a function of the Fermi energy at $\hbar\omega = 2\epsilon_F$ in Bi$_2$Te$_3$ and the red dashed curve shows quadratic fitting.}
	\label{fig:peak}
\end{figure}

The RPE strengthens with $\varepsilon_F$ (as shown in Fig.~\ref{fig:peak}), with the degree of warping ($\lambda$) and the degree of asymmetry of the Fermi surface ($M$). The effect is correspondingly $\propto M$, and, at small $\lambda$ (or small densities), it is $\propto \lambda^2$. Increasing $\lambda$ distorts the Fermi contour from its original circular shape by a larger amount, increasing the current. Conversely, the effect vanishes as $\lambda \rightarrow 0$: as expected, trivially shifting the origin of the Dirac cone by an in-plane magnetization cannot generate a current without the presence of hexagonal warping. Likewise, since the effect is driven by Kramers degeneracy breaking, increasing the asymmetry of the Fermi surface leads to a larger peak. Increasing $\varepsilon_F$ and/or the momentum relaxation time results in a larger Fermi surface displacement. For perfect particle-hole symmetry the height and width of the Lorentzian are controlled only by $\tau$, such that in high mobility systems the peak becomes sharper and taller, and can increase by orders of magnitude. Such an enhancement could be achieved by hybridizing TIs with graphene \cite{Jin_PRB2013, RV_PRB2017, Zhang_SR2018, Song_NL2018, Huang_NLR_PRL2017}. For a small degree of particle-hole asymmetry our conclusions continue to hold, provided the asymmetry does not exceed $\hbar/\tau$.

The TI layer should be as thin as possible so as to enable a strong proximity effect. Strictly speaking, our model applies to films thicker than 3nm, in which there is no tunneling between the top and bottom surfaces \cite{Shan_NJP2010, Weizhe_PRB2014}. However, the effect will be very strong even in thinner films, and our model is still approximately applicable since $\varepsilon_F$ is much larger than the interlayer tunneling strength. If, instead of the ferromagnet, an in-plane magnetic field is used to break time reversal symmetry, in a geometry very similar to Ref.\cite{Yang_NatPhy2018}, the effect will be observable but relatively small due to the inherent smallness of the Bohr magneton. We expect a strong RPE effect in Bi$_{2-x}$Mn$_x$Te$_3$ synthesized recently\cite{Lee_PRB2014, Ruzicka_NJP2015, Carva_PRB2016}, as well as in transition metal dichalcogenides\cite{Kormnyos_2DM2015, xiao_PRL2012}. 

In summary, we have developed the general formalism describing the second order optical response and identified a new, sizable extrinsic contribution to the current which we term RPE. It corresponds to a resonance in the DC photocurrent at $\hbar\omega = 2\epsilon_F$ with a height and width determined by the relaxation time scale. The theory will be extended in a future publication to second harmonic generation, circularly polarized light and transition metal dichalcogenides, whose response is complicated by the finite mass and valley degree of freedom.

\acknowledgments

The authors thank Naoto Nagaosa, Inti Sodemann, Michael Fuhrer, Mark Edmonds, Semonti Bhattacharya, Lan Wang, Elaine Li, Shuyun Zhou, Jimin Zhao, Yongqing Li, Shuichi Murakami, Giovanni Vignale, Steven S.-L Zhang, Yuli Lyanda-Geller, and Stefano Chesi for enlightening discussions. This research is supported by the Australian Research Council Centre of Excellence in Future Low-Energy Electronics Technologies (project CE170100039) and funded by the Australian Government. PB acknowledges the Chinese Postdocs Science Foundation grant No. 2019M650461 and NSAF China grant No. U1930402 for financial support. This research was partially supported by the National Science Foundation through the Center for Dynamics and Control of Materials: an NSF MRSEC under Cooperative Agreement No. DMR-1720595.


%

\appendix

\begin{widetext}

\section{Generalization of Kinetic equation of density matrix}
We begin with the quantum Liouville equation for the density matrix $\langle \rho \rangle$ averaged over impurity configurations, within eigenstate $\vert m,\bm{k} \rangle = e^{i\bm{k}\cdot \bm{r}} \vert u_{\bm{k}}^{m}\rangle$ representation\cite{winkler_book,culcer_17} having $\vert u_{\bm{k}}^{m}\rangle$ a periodic part of the Bloch function with $m$ as band index, ${\bm k}$ as wave vector and ${\bm r}$ as position vector, 
\begin{equation}
\frac{d\langle\rho\rangle}{dt} + \frac{i}{\hbar}[H_0, \langle \rho\rangle] + J[\langle \rho \rangle] = -\frac{i}{\hbar}[H_{E},\langle\rho \rangle].
\label{QKES}
\end{equation}
Here $H_0$ is the band Hamiltonian, $H_E(t) = -e\bf{E}(t)\cdot \hat{r}$ is a perturbation due to a time dependent and spatially inhomogeneous external electric field and $J(\langle \rho \rangle)$ is the scattering term which takes the form with in the Born approximation
\begin{equation}
J(\bkt{\rho}) = \frac{1}{\hbar^2} \, \int^\infty_0 dt' \, \bkt{[U, [e^{-\frac{iH_0t'}{\hbar}}U e^{\frac{iH_0t'}{\hbar}}, \bkt{\rho(t)}]]},
\label{JScatt}
\end{equation} 
having $U$ a random disorder potential.

In the powers of electric field, the density matrix can be expanded alike
\begin{equation}
\langle\rho\rangle = \langle\rho\rangle^{(0)} + \langle\rho\rangle^{(1)} + \langle\rho\rangle^{(2)} + ...,
\end{equation}
where $\langle \rho \rangle^{(0)}$ is an equilibrium density matrix and $\langle \rho \rangle^{(1)}$, $\langle \rho \rangle^{(2)}$,... are corrections to  $\langle \rho \rangle^{(0)}$ to first, second and so on order in the electric field. Using this, the quantum kinetic equation (\ref{QKES}) can be written in a general way
\begin{equation}
\frac{d\langle\rho\rangle^{(n)}}{dt} + \frac{i}{\hbar}[H_0, \langle \rho\rangle^{(n)}] + J[\langle \rho \rangle^{(n)}] = -\frac{i}{\hbar}[H_{E},\langle\rho \rangle^{(n-1)}].
\label{GQKE}
\end{equation}
Further the right hand side of the above equation can be simplified in a manner
\begin{equation}
\ba
-\frac{i}{\hbar}\langle m,{\bm k}|[H_{E}(t),\langle\rho\rangle^{(n-1)}]|m',{\bm k}\rangle&\dps=\frac{e{\bm E(t)}}{\hbar}\cdot \Big\{\pd{\langle\rho\rangle^{(n-1)}}{{\bm k}}-i [\mathcal{R}_{\bm k}, \langle\rho\rangle^{(n-1)}]\Big\}\\[3ex]
&\dps =\frac{e{\bm E}(t)}{\hbar} \cdot \frac{D \langle\rho\rangle^{(n - 1)}} {D {\bm k}}.
\ea
\end{equation}
Here $\frac{D\langle\rho \rangle}{D \bm{k}} = \pd{\langle\rho \rangle}{{\bm k}} - i[\mathcal{R}_{\bm k}, \langle\rho \rangle ]$ is the covariant derivative with respect to wave vector $\bm{k}$ and $\mathcal{\bm R}_{\bm k}^{mm'} = \langle u_{\bm k}^{m}\vert i \frac{\partial u_{\bm k'}^{m'}}{\partial {\bm k}}\rangle$ is the momentum space Berry connection.

To solve the kinetic equation, we decompose the density matrix into two parts on the basis of band index
\begin{equation}
\ba
\langle\rho\rangle^{(n)}&\dps=f_{\text{d}}^{(n)}+f_{\text{od}}^{(n)},
\ea
\end{equation}
where $f_\text{d}$ is diagonal matrix in band index and $f_{\text{od}}$ is off-diagonal. Using this decomposition, Eq.~(\ref{GQKE}) can be segregated into two coupled equations of form
\begin{equation}
\frac{df^{(n)}_{\text{d},{\bm k}}}{dt} + J_{\text{d}}[f^{(n)}_{\text{d},{\bm k}}] = \frac{e{\bm E}(t)}{\hbar} \cdot \frac{D f^{(n - 1)}_{\text{d},{\bm k}}} {D {\bm k}} - J_{\text{d}}[f^{(n)}_{\text{od},{\bm k}}],
\label{GDQKE}
\end{equation}
\begin{equation}
\frac{df^{(n)}_{\text{od},{\bm k}}}{dt} + \frac{i}{\hbar} \, [H_{0{\bm k}}, f^{(n)}_{\text{od},{\bm k}}] + J_{\text{od}}[f^{(n)}_{\text{od},{\bm k}}]  = \frac{e{\bm E}(t)}{\hbar} \cdot \frac{D f^{(n - 1)}_{\text{od},{\bm k}}} {D {\bm k}} - J_{\text{od}}[f^{(n)}_{\text{d},{\bm k}}].
\label{GODQKE}
\end{equation}
Here, we use the decomposition of the scattering term 
\begin{equation}
\ba
J[\langle\rho\rangle^{(n)}] &\dps= J_{\text{d}}[\langle\rho\rangle^{(n)}] + J_{\text{od}}[\langle\rho\rangle^{(n)}],\\[2ex]
J_{\text{d}}[\langle\rho\rangle^{(n)}] &\dps= J_{\text{d}}[f_{\text{d},{\bm k}}^{(n)}] + J_{\text{d}}[f_{\text{od},{\bm k}}^{(n)}]; \quad J_{\text{od}}[\langle\rho\rangle^{(n)}]= J_{\text{od}}[f_{\text{d},{\bm k}}^{(n)}] + J_{\text{od}}[f_{\text{od},{\bm k}}^{(n)}],
\ea
\end{equation}
having general form for $J[\langle \rho \rangle]_{\bm k}$ on solving Eq.~\ref{JScatt}
\begin{equation}
\ba
J[\langle \rho \rangle]_{\bm k}^{mm'''} &\dps= \frac{\pi}{\hbar} \sum_{m', m''{\bm k'}} \bigg\{ \langle U_{{\bm k}{\bm k'}}^{mm'} U_{{\bm k'}{\bm k}}^{m'm''} \rangle \langle \rho \rangle_{\bm k}^{m''m'''}\delta(\varepsilon_{m'{\bm k'}} -\varepsilon_{m''{\bm k}} ) + \langle U_{{\bm k}{\bm k'}}^{m'm''} U_{{\bm k'}{\bm k}}^{m''m'''} \rangle \langle \rho \rangle_{\bm k}^{mm'}\delta(\varepsilon_{m'{\bm k}} -\varepsilon_{m''{\bm k'}} ) \\[2ex]
&\dps  - \langle U_{{\bm k}{\bm k'}}^{mm'} U_{{\bm k'}{\bm k}}^{m''m'''} \rangle \langle \rho \rangle_{\bm k'}^{m'm''}\delta(\varepsilon_{m''{\bm k'}} -\varepsilon_{m'''{\bm k}} ) - \langle U_{{\bm k}{\bm k'}}^{mm'} U_{{\bm k'}{\bm k}}^{m''m'''} \rangle \langle \rho \rangle_{\bm k'}^{m'm''}\delta(\varepsilon_{m{\bm k}} -\varepsilon_{m'{\bm k'}} )\bigg\}.
\ea
\end{equation}
Now, we will solve the quantum Liouville equations for diagonal and off-diagonal parts of density matrices upto second order in the electric field.

\subsection{First order density matrix}
For the first order density matrix ($n=1$), the kinetic equation (Eq.~\ref{GDQKE}) for the diagonal part of density matrix is
\begin{equation}
\frac{df^{(1)}_{\text{d},{\bm k}}}{dt} + J_{\text{d}}[f^{(1)}_{\text{d},{\bm k}}] = \frac{e{\bm E}(t)}{\hbar} \cdot \pd{ f^{(0)}_{\text{d},{\bm k}}}{{\bm k}} - J_{\text{d}}[f^{(1)}_{\text{od},{\bm k}}],
\end{equation}

\begin{equation}
\frac{d f^{(1)}_{\text{d},{\bm k}}}{dt}+\frac{f^{(1)}_{\text{d},{\bm k}}}{\tau}=\frac{e{\bm E}(t)}{\hbar} \cdot \pd{ f^{(0)}_{\text{d},{\bm k}}}{{\bm k}} - J_{\text{d}}[f^{(1)}_{\text{od},{\bm k}}],
\label{DE}
\end{equation}
where $f_{\text{d},{\bm k}}^{(0)} \equiv f_{m,{\bm k}}^{(0)} $ is an equilibrium Fermi-Dirac distribution function and $\tau$ is the momentum relaxation time. 

With the integrating factor, we have
\begin{equation}
f^{(1)}_{\text{d},{\bm k}}(t)  =\int^t_{-\infty} dt' e^{-\frac{t-t'}{\tau}} \bigg\{\frac{e{\bm E(t')}}{\hbar} \cdot\pd{f^{(0)}_{\text{d},{\bm k}}}{{\bm k}} - J_{\text{d}}[f^{(1)}_{\text{od},{\bm k}}(t')] \bigg\}.
\end{equation}
On considering the external electric field of a form $\bm{E}(t) = \bm{E} \cos\omega t$, the above equation becomes
\begin{equation}
\ba
f^{(1)}_{\text{d},{\bm k}}(t) &\dps =\int^t_{-\infty} dt' e^{-\frac{t-t'}{\tau}}\bigg\{\frac{e{\bm E}}{\hbar}  \bigg(\frac{e^{i\omega t'}+e^{-i\omega t'}}{2}\bigg)\cdot\pd{f^{(0)}_{\text{d},{\bm k}}}{{\bm k}}- J_{\text{d}}[f^{(1)}_{\text{od},{\bm k}}(t')] \bigg\}\\[3ex]
&\dps =\frac{e{\bm E}}{2\hbar}\cdot \pd{f^{(0)}_{\text{d},{\bm k}}}{{\bm k}}\bigg(\frac{e^{i\omega t}}{1/\tau+i\omega}+\frac{e^{-i\omega t}}{1/\tau-i\omega}\bigg)- \int_{-\infty}^{t} dt' e^{-(t-t')/\tau} J_\text{d}[f_{\text{od},{\bm k}}^{(1)}(t')].
\label{FOD}
\ea
\end{equation}

The kinetic equation for first-order off-diagonal part of density matrix has a form
\begin{equation}
\frac{df^{(1)}_{\text{od},{\bm k}}}{dt} + \frac{i}{\hbar} \, [H_{0{\bm k}}, f^{(1)}_{\text{od},{\bm k}}] + J_{\text{od}}[f^{(1)}_{\text{od},{\bm k}}]  = \frac{e{\bm E}(t)}{\hbar} \cdot \frac{D f^{(0)}_{\text{od},{\bm k}}} {D {\bm k}} - J_{\text{od}}[f^{(1)}_{\text{d},{\bm k}}].
\end{equation}
\begin{equation}
\frac{df^{(1)}_{\text{od},{\bm k}}}{dt} + \frac{i}{\hbar} \, [H_{0{\bm k}}, f^{(1)}_{\text{od},{\bm k}}] + \frac{f^{(1)}_{\text{od},{\bm k}}}{\tau}  = \frac{e{\bm E}(t)}{\hbar} \cdot \frac{D f^{(0)}_{\text{od},{\bm k}}} {D {\bm k}} - J_{\text{od}}[f^{(1)}_{\text{d},{\bm k}}],
\end{equation}
and the solution is
\begin{equation}
\ba
f^{(1)}_{\text{od},{\bm k}}(t)&\dps =\int^{t}_{-\infty} dt' e^{-\frac{t-t'}{\tau}}e^{-\frac{i\varepsilon_{m,{\bm k}}(t-t')}{\hbar}}\Bigg[-\frac{e{\bm E}(t')}{\hbar}\cdot \Big\{i \mathcal{R}^{mm'}_{\bm k}[f_0(\varepsilon_{m,{\bm k}})-f_0(\varepsilon_{m',\bm k})]\Big\}-J_{\text{od}}[f^{(1)}_{\text{d},{\bm k}}(t')]\Bigg]e^{\frac{i\varepsilon_{m'{\bm k}}(t-t')}{\hbar}}\\[3ex]
&\dps =\int^{t}_{-\infty} dt' e^{-(t- t')/\tau}e^{-\frac{i\varepsilon_{m,{\bm k}}(t-t')}{\hbar}}\Bigg[-\frac{e{\bm E}}{\hbar}\bigg(\frac{e^{i\omega t'}+e^{-i\omega t'}}{2}\bigg)\cdot \Big\{i \mathcal{R}^{mm'}_{\bm k}[f_0(\varepsilon_{m,{\bm k}})-f_0(\varepsilon_{m'{\bm k}})]\Big\}\\[2ex]
&\dps \quad\quad\quad\quad  -J_{\text{od}}[f^{(1)}_{\text{d},{\bm k}}(t')]\Bigg]e^{\frac{i\varepsilon_{m'{\bm k}}(t-t')}{\hbar}}.
\ea
\end{equation}
\begin{equation}
\ba
f^{(1)}_{\text{od},{\bm k}}(t)
&\dps=\sum_{l=\pm}\frac{e \bm{E}}{2}\cdot \frac{e^{il \omega t}\mathcal{R}^{mm'}_{\bm k}[f_0(\varepsilon_{m,{\bm k}})-f_0(\varepsilon_{m'{\bm k}})]}{(l\hbar\omega+\varepsilon_{m,{\bm k}}-\varepsilon_{m'{\bm k}})-i\frac{\hbar}{\tau}}-\int_{-\infty}^t dt' e^{-\frac{t-t'}{\tau}} e^{-\frac{i \varepsilon_{m,{\bm k}}(t-t')}{\hbar}} J_{\text{od}}[f_{\text{d},{\bm k}}^{(1)}(t')]e^{\frac{i \varepsilon_{m',{\bm k}}(t-t')}{\hbar}}.
\label{FDOD}
\ea
\end{equation}

\subsection{Second-order density matrix}
For the second-order case, we set $n=2$ in Eq.~(\ref{GQKE}). The diagonal part of the equation is
\begin{equation}
\frac{df^{(2)}_{\text{d},{\bm k}}}{dt} + J_{\text{d}}[f^{(2)}_{\text{d},{\bm k}}] = \frac{e{\bm E}(t)}{\hbar} \cdot \pd{ f^{(1)}_{\text{d},{\bm k}}}{{\bm k}} - J_{\text{d}}[f^{(2)}_{\text{od},{\bm k}}],
\end{equation}

\begin{equation}
\frac{d f^{(2)}_{\text{d},{\bm k}}}{dt}+\frac{f^{(2)}_{d,{\bm k}}}{\tau}=\frac{e{\bm E}(t)}{\hbar} \cdot \pd{ f^{(1)}_{\text{d},{\bm k}}}{{\bm k}} - J_{\text{d}}[f^{(2)}_{\text{od},{\bm k}}].
\label{SDE}
\end{equation}

On solving linear differential equation, we have
\begin{equation}
f^{(2)}_{\text{d},{\bm k}}(t)  =\int^t_{-\infty} dt' e^{-\frac{t-t'}{\tau}} \bigg\{\frac{e{\bm E(t')}}{\hbar} \cdot\pd{f^{(1)}_{d,{\bm k}}(t')}{{\bm k}} - J_{\text{d}}[f^{(2)}_{\text{od},{\bm k}}(t')] \bigg\}.
\end{equation}

\begin{equation}
\ba
f^{(2)}_{\text{d},{\bm k}}(t) &\dps =\int^t_{-\infty} dt' e^{-\frac{t-t'}{\tau}}\bigg\{\frac{e{\bm E}}{\hbar}  \bigg(\frac{e^{i\omega t'}+e^{-i\omega t'}}{2}\bigg)\cdot\pd{f^{(1)}_{\text{d},{\bm k}}(t')}{{\bm k}}- J_{\text{d}}[f^{(2)}_{\text{od},{\bm k}}(t')]^{} \bigg\}\\[3ex]
\ea
\end{equation}

Further substitution of $f^{(1)}_{\text{d},{\bm k}}(t)$ (Eq.~\ref{FOD}) in the above equation yields
\begin{equation}
\ba
f^{(2)}_{\text{d},{\bm k}}(t) &\dps=\int^t_{-\infty} dt' e^{-\frac{t-t'}{\tau}}\frac{e{\bm E}}{\hbar}  \bigg(\frac{e^{i\omega t'}+e^{-i\omega t'}}{2}\bigg)\cdot\pd{}{{\bm k}}\bigg[  \frac{e{\bm E}}{2\hbar}\cdot \pd{f^{(0)}_{d,{\bm k}}}{{\bm k}}\bigg(\frac{e^{i\omega t'}}{1/\tau+i\omega}+\frac{e^{-i\omega t'}}{1/\tau-i\omega}\bigg)\bigg]\\[3ex]
&\dps - \int^t_{-\infty} dt' e^{-\frac{t-t'}{\tau}}\bigg\{\frac{e{\bm E}}{\hbar}  \bigg(\frac{e^{i\omega t'}+e^{-i\omega t'}}{2}\bigg)\cdot\pd{}{{\bm k}}\bigg[\int_{-\infty}^{t'} dt'' e^{-\frac{t'-t''}{\tau}} J_\text{d}[f_{\text{od},{\bm k}}^{(1)}(t'')]\bigg]\bigg\}\\[3ex]
&\dps -\int^t_{-\infty} dt' e^{-\frac{t-t'}{\tau}} J_{\text{d}}[f^{(2)}_{\text{od},{\bm k}}(t')]. 
\ea
\end{equation}

\begin{equation}
\ba
f^{(2)}_{\text{d},{\bm k}}(t) &\dps =\int^t_{-\infty} dt' e^{-\frac{t-t'}{\tau}}\frac{e{\bm E}}{2\hbar}\sum_{l=\pm}\Big(\frac{e^{i2l\omega t'}}{1/\tau+il\omega} + \frac{1}{1/\tau+il\omega}\Big)  \cdot\pd{}{{\bm k}}\bigg[  \frac{e{\bm E}}{2\hbar}\cdot \pd{f^{(0)}_{d,{\bm k}}}{{\bm k}}\bigg]\\[3ex]
&\dps - \int^t_{-\infty} dt' e^{-\frac{t-t'}{\tau}}\bigg\{\frac{e{\bm E}}{\hbar}  \big[\frac{e^{i\omega t'}+e^{-i\omega t'}}{2}\big]\cdot\pd{}{{\bm k}}\bigg[\int_{-\infty}^{t'} dt'' e^{-\frac{t'-t''}{\tau}} J_\text{d}[f_{\text{od},{\bm k}}^{(1)}(t'')]\bigg]\bigg\}\\[3ex]
&\dps -\int^t_{-\infty} dt' e^{-\frac{t-t'}{\tau}} J_{\text{d}}[f^{(2)}_{\text{od},{\bm k}}(t')]. 
\ea
\end{equation}

\begin{equation}
\ba
f^{(2)}_{d,{\bm k}}(t) &\dps = \frac{e^2 \bm{E}}{4\hbar^2}\cdot \pd{}{\bm{k}} \bigg(\bm{E}\cdot\pd{ f_{0 k}}{\bm{k}}\bigg) \bigg\{\sum_{l=\pm}\frac{e^{i 2l\omega t}}{(il\omega + \tau^{-1})(i2l\omega + \tau^{-1})}  + \frac{2}{\omega^2 + \tau^{-2}} \bigg\}\\[3ex]
&\dps - \int^t_{-\infty} dt' e^{-\frac{t-t'}{\tau}}\bigg\{\frac{e{\bm E}}{\hbar}  \bigg(\frac{e^{i\omega t'}+e^{-i\omega t'}}{2}\bigg)\cdot\pd{}{{\bm k}}\bigg[\int_{-\infty}^{t'} dt'' e^{-\frac{t'-t''}{\tau}} J_\text{d}[f_{\text{od},{\bm k}}^{(1)}(t'')]\bigg]\bigg\}\\[3ex]
&\dps - \int_{-\infty}^{t} dt' e^{-\frac{t-t'}{\tau}} J_{\text{d}}[f_{\text{od},{\bm k}}^{(2)}(t')].
\ea
\end{equation}
This is a general solution of kinetic equation for second-order diagonal density matrix.

The kinetic equation for second-order off-diagonal matrix $f^{(2)}_{od,{\bm k}}$ is 
\begin{equation}
\frac{df^{(2)}_{\text{od},{\bm k}}}{dt} + \frac{i}{\hbar} \, [H_{0{\bm k}}, f^{(2)}_{\text{od},{\bm k}}] + J_{\text{od}}[f^{(2)}_{\text{od},{\bm k}}]  = \frac{e{\bm E}(t)}{\hbar} \cdot \frac{D f^{(1)}_{\text{od},{\bm k}}} {D {\bm k}} - J_{\text{od}}[f^{(2)}_{\text{d},{\bm k}}].
\end{equation}
\begin{equation}
\frac{df^{(2)}_{\text{od},{\bm k}}}{dt} + \frac{i}{\hbar} \, [H_{0{\bm k}}, f^{(2)}_{\text{od},{\bm k}}] + \frac{f^{(2)}_{\text{od},{\bm k}}}{\tau}  = \frac{e{\bm E}(t)}{\hbar} \cdot \frac{D f^{(1)}_{\text{od},{\bm k}}} {D {\bm k}} - J_{\text{od}}[f^{(2)}_{\text{d},{\bm k}}],
\end{equation}
and the solution is
\begin{equation}
\ba
f^{(2)}_{\text{od},{\bm k}}(t)&\dps =\int^{t}_{-\infty} dt' e^{-\frac{t- t'}{\tau}}e^{-\frac{i\varepsilon_{m,{\bm k}}(t-t')}{\hbar}}\bigg[-\frac{e{\bm E}(t')}{\hbar}\cdot \bigg\{\pd{f^{(1)}_{od,{\bm k}}(t')}{{\bm k}}-i [\mathcal{R}_{\bm k}, \langle\rho\rangle^{(1)}]\bigg\}-J_{\text{od}}[f^{(2)}_{\text{d},{\bm k}}(t')]\bigg]e^{\frac{i\varepsilon_{m'{\bm k}}(t-t')}{\hbar}}\\[3ex]

&\dps =\int^{t}_{-\infty} dt' e^{-\frac{t- t'}{\tau}}e^{-\frac{i\varepsilon_{m,{\bm k}}(t-t')}{\hbar}}\bigg[-\frac{e{\bm E}}{\hbar}\bigg(\frac{e^{i\omega t'}+e^{-i\omega t'}}{2}\bigg)\cdot \Bigg\{\pd{f^{(1)}_{od,{\bm k}}(t')}{{\bm k}}-i [\mathcal{R}_{\bm k}, \langle\rho\rangle^{(1)}]\Bigg\} -J_{\text{od}}[f^{(2)}_{\text{d},{\bm k}}(t')]\Bigg]e^{\frac{i\varepsilon_{m'{\bm k}}(t-t')}{\hbar}}.
\ea
\label{SOOD1}
\end{equation}
The commutator $[\mathcal{R}_{\bm k}, \langle\rho\rangle^{(1)}]^{mm'}$ can be simplified as
\begin{equation}
\ba
[\mathcal{R}_{\bm k}, \langle\rho\rangle^{(1)}]^{mm'}
&\dps=\sum_{m''}\langle m|\mathcal{R}_{\bm k}|m''\rangle\langle m''|\langle\rho\rangle^{(1)}|m'\rangle-\langle m|\langle\rho^{(1)}|m''\rangle \langle m''|\mathcal{R}_{\bm k}|m'\rangle\\[3ex]
&\dps = f_{\text{od},{\bm k}}^{{(1)},mm'}(\mathcal{R}_{\bm k}^{mm} - \mathcal{R}_{\bm k}^{m'm'}) + \mathcal{R}_{\bm k}^{mm'}(f_{\text{d},{\bm k}}^{{(1)},m'm'}-f_{\text{d},{\bm k}}^{{(1)},mm})
\ea
\end{equation}
With the help of this commutator and Eqs.~(\ref{FOD}) and (\ref{FDOD}), Eq.~(\ref{SOOD1}) takes form
\begin{equation}
\ba
f^{(2)}_{\text{od},{\bm k}}(t)&\dps =\int^{t}_{-\infty} dt' e^{-\frac{t- t'}{\tau}}e^{-\frac{i\varepsilon_{m,{\bm k}}(t-t')}{\hbar}}\Bigg[-\frac{e{\bm E}}{\hbar}\bigg(\frac{e^{i\omega t'}+e^{-i\omega t'}}{2}\bigg)\cdot \bigg\{ \pd{}{{\bm k}} \Big( F_{1{\bm k}}(t')+G_{1{\bm k}}(t')\Big)\\[3ex]

&\dps-i \Big(F_{1{\bm k}}(t')- G_{1{\bm k}}(t')\Big)(\mathcal{R}_{\bm k}^{mm} - \mathcal{R}_{\bm k}^{m'm'}) -i \mathcal{R}_{\bm k}^{mm'}\Big( F_{2,{\bm k}}(t')+G_{2,{\bm k}}(t')\Big)\bigg\}\Bigg]e^{\frac{i\varepsilon_{m'{\bm k}}(t-t')}{\hbar}}\\[3ex]

&\dps   -\int^{t}_{-\infty} dt' e^{-\frac{t- t'}{\tau}}e^{-\frac{i\varepsilon_{m,{\bm k}}(t-t')}{\hbar}}J_{\text{od}}[f^{(2)}_{\text{d},{\bm k}}(t')]e^{\frac{i\varepsilon_{m'{\bm k}}(t-t')}{\hbar}}.
\ea
\end{equation}
Here
\begin{equation}
\ba
F_{1{\bm k}}(t) &\dps =\sum_{l=\pm}\frac{e^{il \omega t}e\bm{E}\cdot\mathcal{R}^{mm'}_{\bm k}[f_0(\varepsilon_{m,{\bm k}})-f_0(\varepsilon_{m'{\bm k}})]}{2(l\hbar\omega+\varepsilon_{m,{\bm k}}-\varepsilon_{m'{\bm k}})-i\frac{\hbar}{\tau}}; \quad G_{1{\bm k}}(t) =\int_{-\infty}^t dt' e^{-\frac{t-t'}{\tau}} e^{-\frac{i \varepsilon_{m,{\bm k}}(t-t')}{\hbar}} J_{\text{od}}[f_{\text{d},{\bm k}}^{(1)}(t')]e^{\frac{i \varepsilon_{m',{\bm k}}(t-t')}{\hbar}},\\[3ex]

F_{2,{\bm k}}(t) &\dps = \sum_{l=\pm}\frac{e^{il\omega t}e\bm{E}\cdot}{2(\hbar/\tau+il\hbar\omega)} \pd{}{{\bm k}}(f_{\text{d},{\bm k}}^{{(0)},m'm'}-f_{\text{d},{\bm k}}^{{(0)},mm}); \quad G_{2,{\bm k}}(t) = \int_{-\infty}^{t} dt' e^{-\frac{t-t'}{\tau}} (J_\text{d}[f_{\text{od},{\bm k}}^{(1)}(t')]^{m'm'}-J_\text{d}[f_{\text{od},{\bm k}}^{(1)}(t')]^{mm})
\ea
\end{equation}
On further simplifications, we have

\begin{equation}
\ba
f^{(2)}_{od,{\bm k}}(t)  &\dps =  \frac{e \bm{E}}{2}\cdot \sum_{l=\pm}\bigg\{\frac{e^{i2l\omega t}}{i(2l\hbar\omega +\varepsilon_{m{\bm k}} -\varepsilon_{m'{\bm k}})+\hbar\tau^{-1}} + \frac{1}{i(\varepsilon_{m,{\bm k}} -\varepsilon_{m'{\bm k}}) +\hbar\tau^{-1} }  \bigg\}\\[2ex]
&\dps \times \bigg(\pd{F_{1k}}{\bm{k}} -iF_{1k}(\mathcal{R}_{\bm k}^{mm} - \mathcal{R}_{\bm k}^{m'm'})-i\mathcal{R}_{\bm k}^{mm'}F_{2,{\bm k}} \bigg)\\[3ex]

&\dps -\int^{t}_{-\infty} dt' e^{-\frac{t- t'}{\tau}}e^{-\frac{i\varepsilon_{m,{\bm k}}(t-t')}{\hbar}}\Bigg[\frac{e{\bm E}}{\hbar}\bigg(\frac{e^{i\omega t'}+e^{-i\omega t'}}{2}\bigg)\cdot \bigg( \pd{}{{\bm k}} G_{1{\bm k}}(t')+iG_{1{\bm k}}(t')(\mathcal{R}_{\bm k}^{mm} - \mathcal{R}_{\bm k}^{m'm'})  \bigg)\bigg]e^{\frac{i\varepsilon_{m'{\bm k}}(t-t')}{\hbar}}\\[3ex]

&\dps +\int^{t}_{-\infty} dt' e^{-\frac{t- t'}{\tau}}e^{-\frac{i\varepsilon_{m,{\bm k}}(t-t')}{\hbar}}\Bigg[\frac{e{\bm E}}{\hbar}\bigg(\frac{e^{i\omega t'}+e^{-i\omega t'}}{2}\bigg)\cdot i\mathcal{R}_{\bm k}^{mm'} G_{2,{\bm k}}(t')- J_{\text{od}}[f_{\text{d},{\bm k}}^{(2)}(t')]\Bigg]e^{\frac{i\varepsilon_{m'{\bm k}}(t-t')}{\hbar}}.
\ea
\end{equation}

This is a general solution of kinetic equation for second-order off-diagonal density matrix. 

In the present work, we are mainly keen in time independent term contribution which leads to shift current. For absorptive part, the diagonal and off-diagonal density matrices can be read as
\begin{equation}
\ba
f^{(2)}_{\text{d},{\bm k}} &\dps = \frac{e^2 \bm{E}}{4}\cdot \pd{}{\bm{k}} \bigg(\bm{E}\cdot\pd{ f_{\bm k}^{0}}{\bm{k}}\bigg)  \frac{2}{\omega^2 + \tau^{-2}}  - \tau J_{\text d}[f_{\text{od},{\bm k}}^{(2)}],
\label{f_diagonal}
\ea
\end{equation}
\begin{equation}
\ba
f^{(2)}_{\text{od},{\bm k}}  &\dps =   \frac{e \bm{E}}{2}\cdot \bigg\{\frac{1}{i(\varepsilon_{m,{\bm k}} -\varepsilon_{m'{\bm k}}) +\frac{\hbar}{\tau} }  \bigg\}\bigg(\pd{F_{1{\bm k}}}{\bm{k}} -iF_{1{\bm k}}(\mathcal{R}_{\bm k}^{mm} - \mathcal{R}_{\bm k}^{m'm'})-i\mathcal{R}_{\bm k}^{mm'}F_{2,{\bm k}} \bigg)- \frac{J_{\text{od}}[f_{\text{d},{\bm k}}^{(2)}]}{i(\varepsilon_{m{\bm k}} -\varepsilon_{m'{\bm k}}) +\frac{\hbar}{\tau}}.
\ea
\end{equation}
After substituting the expressions for $F_{1,{\bm k}}$ and $F_{2,{\bm k}}$ and performing algebraic calculations, we obtain
\begin{equation}
\ba
f^{(2)}_{\text{od},{\bm k}}&\dps  = \frac{e^2 {\bm E}}{4}\cdot \bigg[-i{\bm E}\cdot \bigg(\frac{\mathcal{R}_{{\bm k}}^{mm'}}{\hbar\omega}\pd{f_{{\bm k}}^{0,mm'}}{{\bm k}} + \frac{f_{\bm k}^{0,mm'}}{\hbar\omega}\pd{ \mathcal{R}_{{\bm k}}^{mm'}}{{\bm k}}+\frac{\mathcal{R}_{{\bm k}}^{mm'}f_{{\bm k}}^{0,mm'}}{(\hbar\omega)^2} \pd{(\varepsilon_{m{\bm k}}-\varepsilon_{m'{\bm k}})}{{\bm k}}+\frac{(\mathcal{R}_{{\bm k}}^{mm}-\mathcal{R}_{{\bm k}}^{m'm'})\mathcal{R}_{{\bm k}}^{mm'}f_{{\bm k}}^{0,mm'}}{\hbar\omega }\bigg)\\[3ex]
&\dps \times \bigg(\frac{-1}{\varepsilon_{m{\bm k}}-\varepsilon_{m'{\bm k}} -i\frac{\hbar}{\tau}} +\frac{1}{(\varepsilon_{m{\bm k}}-\varepsilon_{m'{\bm k}}-\hbar\omega)-i\frac{\hbar}{\tau}}  \bigg)\\[3ex]
&\dps-\frac{i{\bm E}\cdot \mathcal{R}_{{\bm k}}^{mm'}f_{{\bm k}}^{0,mm'}}{\hbar\omega} \pd{(\varepsilon_{m{\bm k}}-\varepsilon_{m'{\bm k}})}{{\bm k}}\frac{1}{\big((\varepsilon_{m{\bm k}}-\varepsilon_{m'{\bm k}}-\hbar\omega)-i\frac{\hbar}{\tau}\big)^2} +\frac{i2{\bm E}\cdot R_{{\bm k}}^{mm'}\tau^{-1}}{\omega^2 + \tau^{-2}}\frac{1}{\varepsilon_{m{\bm k}}-\varepsilon_{m'{\bm k}} -i\frac{\hbar}{\tau}}\pd{f_{{\bm k}}^{0}}{{\bm k}}  \bigg]\\[3ex]
&\dps- \frac{J_{\text{od}}[f_{\text{d},{\bm k}}^{(2)}]}{i(\varepsilon_{m{\bm k}} -\varepsilon_{m'{\bm k}}) +\frac{\hbar}{\tau}},
\label{f_offdiagonal}
\ea
\end{equation}
where we use $f_{{\bm k}}^{0,mm'}=f^0(\varepsilon_{m{\bm k}})-f^0(\varepsilon_{m'{\bm k}})$.
\section{Hamiltonian and other relevant quantities}
The Hamiltonian of a system is
\begin{equation}
\ba
H_{\bm k} &\dps=A(\sigma_xk_y-\sigma_yk_x)+\lambda\sigma_z(k^3_x-3k_xk^2_y)+{\bm \sigma}\cdot {\bm M}, \\[3ex]
&\dps=A k(\sin\theta \sigma_x -\cos\theta \sigma_y)+\lambda k^3\cos3\theta\sigma_z + \sigma_y M \\[3ex]
\ea
\end{equation}
where $A=\alpha\hbar$ is the spin-orbit constant, $\sigma_i$'s are the Pauli matrices, $\theta= \arctan(k_y/k_x)$ is the polar angle of wave-vector ${\bm k}$ and $\bm{M}$ is the magnetization along $y$-direction having $\bm{M}=M \hat{y}$. The eigen values and eigen vectors are

\begin{equation}
\ba
\varepsilon_{k,\pm}&\dps=\pm\sqrt{A^2 k^2 + M^2 + 2A k M \cos\theta + \lambda^2 k^6 \cos^23\theta}\\[3ex]
|u^\pm_{\bm k}\rangle &\dps  =\frac{1}{\sqrt{2}}
\left(
\begin{array}{c}
\mp i\sqrt{1\pm b_k}\frac{(M - A k e^{-i\theta})}{a_k \epsilon_k
}   \\
\sqrt{1\mp b_k}
\end{array}
\right).\\[3ex]
\ea
\end{equation}
Here $m=\pm$ represents the two band indices and $a_k$, $b_k$ are defined as
\begin{equation}
\ba
a_k &\dps =\frac{\sqrt{A^2 k^2 + M^2 + 2A k M \cos\theta}}{\epsilon_k}, \quad b_k=\frac{\lambda k^3\cos3\theta}{\epsilon_k}, \qquad a^2_k+b^2_k=1.\\[3ex]
\ea
\end{equation}
Further the Berry connection part for different band indices combination is
\begin{equation}
\ba
\mathcal{R}^{\pm\pm}_{\bm k}&\dps=\frac{- A B \sin\theta}{2a_k^2 \epsilon_k^2}\bigg[ 1\pm \frac{\lambda k^3 \cos3\theta}{\epsilon_k}\bigg]\hat{{\bm k}} + \frac{A(A k - M \cos\theta))}{2 a_k^2\epsilon_k^2}\bigg[ 1\pm \frac{\lambda k^3 \cos3\theta}{\epsilon_k}\bigg]\hat{{\bm \theta}}\\[3ex]
\ea
\end{equation}

\begin{equation}
\ba
\mathcal{R}^{+-}_{\bm k} &\dps = \frac{i}{4 a_k}\bigg[\frac{2k^2(3M^2 + 2 A^2 k^2)\lambda \cos3\theta - 5A  M \lambda k^3 (\cos 4\theta+\cos 2\theta)}{\epsilon_k^3}   \bigg]\hat{{\bm k}} \\[3ex]
&\dps +\frac{i}{4 a_k}\bigg[ \frac{ \lambda k^2(5 A M k \sin 4\theta + 7AM k \sin 2\theta - 6(M^2+A^2k^2)\sin3\theta)}{\epsilon_k^3}\bigg] \hat{{\bm \theta}}\\[3ex]
&\dps +\frac{1}{2 a_k \epsilon_k^2}\bigg[M \sin\theta \hat{{\bm k}} +(M \cos\theta - A k)\hat{\bm \theta}\bigg]\\[3ex]
\mathcal{R}^{-+}_{\bm k} &\dps  = [\mathcal{R}^{+-}_{\bm k}]^*; \quad  \hat{\bm k}=\hat{\bm x} \cos\theta +\hat{\bm y}\sin\theta,\quad \hat{\bm \theta}=-\hat{\bm x}\sin\theta+\hat{\bm y}\cos\theta.
\ea
\end{equation}

In this work, we consider the disorder as $U(\bm r) = U_0\sum_i\delta({\bm r}-{\bm r}_i)$ and define matrix elements of $U_{{\bm k}{\bm k'}}$ as
\begin{equation}
\ba
&\dps U^{\pm\pm}_{{\bm k}{\bm k}'}=U\langle u^\pm_{\bm k}|u^\pm_{{\bm k}'}\rangle=\frac{U}{2}\Big\{\frac{\sqrt{(1\mp b_k)(1\mp b_{k'})}(M + A k e^{-i\theta})(M +  A k' e^{i\theta'})}{\sqrt{M^2 + A^2k^2 + 2 A M k \cos\theta}\sqrt{M^2 + A^2 k'^2 + 2 A M k' \cos\theta'}}+\sqrt{(1\pm b_k)(1\pm b_{k'})}\Big\},\\[3ex]
&\dps U^{\pm\mp}_{{\bm k}{\bm k}'}=U\langle u^\pm_{\bm k}|u^\mp_{{\bm k}'}\rangle=\frac{U}{2}\Big\{\sqrt{(1+b_k)(1-b_{k'})}\mp\frac{\sqrt{(1-b_k)(1+b_{k'})}(M + A k e^{-i\theta})(M +  A k' e^{i\theta'})}{\sqrt{M^2 + A^2k^2 + 2 A M k \cos\theta}\sqrt{M^2 + A^2 k'^2 + 2 A M k' \cos\theta'}}\Big\}.\\[3ex]
\ea
\end{equation}
Using the definitions of matrix elements, the products of matrix elements which are needed to evaluate scattering term can be written
\begin{equation}
\ba
U_{kk'}^{++}U_{k'k}^{+-} &\dps = \frac{U^2}{4a_k a_{k'} \epsilon_k \epsilon_{k'}} \bigg\{(\sqrt{(1+b_k)(1-b_{k'})} - (M-A ke^{-i\theta})(M-A k' e^{i\theta'})\sqrt{(1-b_k)(1+b_{k'})})\\[3ex]
&\dps  \times (\sqrt{(1-b_k)(1-b_{k'})} + (M-A ke^{i\theta})(M-A k' e^{-i\theta'})\sqrt{(1+b_k)(1+b_{k'})})\bigg\}\\[3ex]

U_{kk'}^{-+}U_{k'k}^{++} &\dps =\frac{U^2}{4a_k a_{k'} \epsilon_k \epsilon_{k'}} \bigg\{(\sqrt{(1+b_k)(1-b_{k'})} - (M-A ke^{i\theta})(M-A k' e^{-i\theta'})\sqrt{(1-b_k)(1+b_{k'})})\\[3ex]
&\dps  \times (\sqrt{(1-b_k)(1-b_{k'})} + (M-A k e^{-i\theta})(M-A k' e^{i\theta'})\sqrt{(1+b_k)(1+b_{k'})})\bigg\}\\[3ex]

U_{kk'}^{+-}U_{k'k}^{--} &\dps =\frac{U^2}{4a_k a_{k'} \epsilon_k \epsilon_{k'}} \bigg\{(\sqrt{(1-b_k)(1+b_{k'})} - (M-A k'e^{-i\theta'})(M-A k e^{i\theta})\sqrt{(1+b_k)(1-b_{k'})})\\[3ex]
&\dps  \times (\sqrt{(1+b_k)(1+b_{k'})} + (M-A k'e^{i\theta'})(M-A k e^{-i\theta})\sqrt{(1-b_k)(1-b_{k'})})\bigg\}\\[3ex]

U_{kk'}^{--}U_{k'k}^{-+} &\dps =\frac{U^2}{4a_k a_{k'} \epsilon_k \epsilon_{k'}} \bigg\{(\sqrt{(1-b_k)(1+b_{k'})} - (M-A ke^{-i\theta})(M-A k' e^{i\theta'})\sqrt{(1+b_k)(1-b_{k'})})\\[3ex]
&\dps  \times (\sqrt{(1+b_k)(1+b_{k'})} + (M-A ke^{i\theta})(M-A k' e^{-i\theta'})\sqrt{(1-b_k)(1-b_{k'})})\bigg\}\\[3ex]

U_{kk'}^{+-}U_{k'k}^{++} &\dps =\frac{U^2}{4a_k a_{k'} \epsilon_k \epsilon_{k'}} \bigg\{(\sqrt{(1-b_k)(1+b_{k'})} - (M-A ke^{i\theta})(M-A k' e^{-i\theta'})\sqrt{(1+b_k)(1-b_{k'})})\\[3ex]
&\dps  \times (\sqrt{(1-b_k)(1-b_{k'})} + (M-A ke^{-i\theta})(M-A k' e^{i\theta'})\sqrt{(1+b_k)(1+b_{k'})})\bigg\}\\[3ex]

U_{kk'}^{++}U_{k'k}^{-+} &\dps =\frac{U^2}{4a_k a_{k'} \epsilon_k \epsilon_{k'}} \bigg\{(\sqrt{(1-b_k)(1+b_{k'})} - (M-A ke^{-i\theta})(M-A k' e^{i\theta'})\sqrt{(1+b_k)(1-b_{k'})})\\[3ex]
&\dps  \times (\sqrt{(1-b_k)(1-b_{k'})} + (M-A ke^{i\theta})(M-A k' e^{-i\theta'})\sqrt{(1+b_k)(1+b_{k'})})\bigg\}\\[3ex]
\ea
\end{equation}

\section{RPE current}

The electric current generated by an electric field is defined as
\begin{equation}
\ba
\bm{j} &\dps= -\frac{e}{\hbar} \int \frac{k dk d\theta}{4\pi^2}\bigg(\pd{\varepsilon_{{\bm k}}}{{\bm k}}\delta_{mm'} - i \mathcal{R}_{{\bm k}}^{mm'} (\varepsilon_{m{\bm k}} - \varepsilon_{m'{\bm k}} )  \bigg)f_{{\bm k}}
\ea
\end{equation}
For second-order case, we replace $f_{{\bm k}} \rightarrow f_{{\bm k}}^{(2)}=f_{\text{d},{\bm k}}^{(2)}+f_{\text{od},{\bm k}}^{(2)}$.
\begin{equation}
\ba
\bm{j}^{(2)} &\dps=-\frac{e}{\hbar} \int \frac{k dk d\theta}{4\pi^2}\bigg(\pd{\varepsilon_{{\bm k}}}{{\bm k}}f_{\text{d},{\bm k}}^{(2)} - i \mathcal{R}_{{\bm k}}^{mm'} (\varepsilon_{m{\bm k}} - \varepsilon_{m'{\bm k}} )f_{\text{od},{\bm k}}^{(2)}  \bigg)
\ea
\end{equation}

Considering $m=+1$ and $m'=-1$ for Dirac case and $\bm{E} \parallel x$, the x-component of the off-diagonal part of the shift current within limit $\omega\tau \gg 1$ will be
\begin{equation}
\ba
j_{x,{\text od}}^{(2)} &\dps=\frac{e}{\hbar} \int \frac{k dk d\theta}{4\pi^2} i\mathcal{R}_{{\bm k}}^{+-} (\varepsilon_{+{\bm k}} - \varepsilon_{-{\bm k}} )f_{\text{od},{\bm k}}^{(2)} \\[3ex]

&\dps= \frac{e^3E_x^2}{16\pi^2\hbar}\int dk d\theta  \bigg[k\bigg(2\vert\mathcal{R}_{k_x}^{+-}\vert^2\pd{f_{0{\bm k}}^{+-}}{k_x} + f_{0{\bm k}}^{+-}\pd{\vert\mathcal{R}_{k_x}^{+-}\vert^2}{k_x}+\frac{\vert\mathcal{R}_{k_x}^{+-}\vert^2f_{0{\bm k}}^{+-}}{\hbar\omega} \pd{2\epsilon_k^+}{k_x}+(\mathcal{R}_{k_x}^{++}-\mathcal{R}_{k_x}^{--})\vert\mathcal{R}_{k_x}^{+-}\vert^2f_{0{\bm k}}^{+-} \bigg)\\[3ex]
&\dps+\cos\theta \frac{\vert\mathcal{R}_{k_x}^{+-}\vert^2f_{0{\bm k}}^{+-}}{\hbar\omega}\bigg] \frac{\frac{\hbar}{\tau}}{(2\epsilon_k^+-\hbar\omega)^2+(\frac{\hbar}{\tau})^2}.
\label{ODC}
\ea
\end{equation}
At low temperature, the first term dominates over other terms due to the presence of the wave vector derivative of Fermi factors. Thus, the expression reduces to
\begin{equation}
\ba
j_{x,{\text od}}^{(2)} &\dps= \frac{e^3E_x^2}{8\pi^2\hbar}\int dk d\theta  k\vert\mathcal{R}_{k_x}^{+-}\vert^2\pd{f_{0{\bm k}}^{+-}}{k_x} \frac{\frac{\hbar}{\tau}}{(2\epsilon_k^+-\hbar\omega)^2+(\frac{\hbar}{\tau})^2}.
\label{ODC2}
\ea
\end{equation}

Similarly, the diagonal part contribution yields
\begin{equation}
\ba
j_{x,{\text d}}^{(2)}  &\dps=-\frac{e}{\hbar} \int \frac{k dk d\theta}{4\pi^2}\pd{\varepsilon_{{\bm k}}}{{\bm k}}f_{\text{d},{\bm k}}^{(2)}\\[3ex]

&\dps = \frac{e}{4\pi^2\hbar} \int k dk d\theta\pd{\varepsilon_{{\bm k}}}{k_x} \tau J_{\text{d}}[f_{\text{od},{\bm k}}^{(2)}].
\label{DC}
\ea
\end{equation}
The total current is the sum of the contributions by diagonal and off-diagonal part of density matrices and the behavior of it is shown in the main paper.

\end{widetext}

\end{document}